\documentclass[twocolumn,amssymb,nobibnotes,superscriptaddress,aps,pra,tightenlines]{revtex4-1}
\usepackage{color,graphicx,amsfonts,amssymb,amsmath,footmisc,fancyhdr,subfigure,braket}
\usepackage[colorlinks=true, linkcolor=red, citecolor=blue, urlcolor=blue]{hyperref}

\begin{document}
\title{Imaginary Gauge-steerable Edge Modes In Non-Hermitian Aubry-Andr\'e-Harper Model}
\author{Yazhuang Miao}
\affiliation{School of Science, Qingdao University of Technology, Qingdao, Shandong, China}
\author{Wei Ding}
\affiliation{School of Science, Qingdao University of Technology, Qingdao, Shandong, China}
\author{Litong Wang}
\affiliation{School of Science, Qingdao University of Technology, Qingdao, Shandong, China}
\author{Xiaolong Zhao}
\email{zhaoxiaolong@qut.edu.cn}
\affiliation{School of Science, Qingdao University of Technology, Qingdao, Shandong, China}
\author{Shengguang Liu}
\email{liushengguang@qut.edu.cn}
\affiliation{School of Science, Qingdao University of Technology, Qingdao, Shandong, China}
\author{Xuexi Yi}
\email{yixx@nenu.edu.cn}
\affiliation{Center for Quantum Sciences and School of Physics, Northeast Normal University, Changchun, Jilin, China}
\date{\today}

\begin{abstract}
We identify steerable exponentially localized in-gap mode in a quasiperiodic non-Hermitian Aubry-Andr\'e-Harper chain 
with a spatially fluctuating, zero-mean imaginary gauge field. Under open boundary conditions, the system is exactly 
related to the Hermitian AAH model by a nonunitary gauge transformation: the OBC spectrum and Lyapunov exponents are 
unchanged, while eigenstates acquire a gauge-dependent envelope. In a parameter region with spectrally isolated in-gap 
boundary modes, we find two exponentially localized in-gap modes with sharply different responses to the imaginary 
gauge field. One remains boundary pinned, but the other is gauge-steerable: it stays exponentially localized while its 
probability maximum shifts as the gauge field is changed, with its eigenenergy unchanged. We further show that weak 
on-site gain, applied at a single site chosen once and then kept fixed, can dynamically prepare this steerable mode 
from a generic bulk wave packet. Changing the gauge field then yields exponentially localized states at different locations.
\end{abstract}

\maketitle

\section{Introduction}

Non-Hermitian Hamiltonians provide an effective description of driven or open quantum systems with gain and loss,
leakage, measurement backaction, or nonreciprocal couplings. Such systems host phenomena absent in Hermitian 
settings, including parity-time symmetry breaking, exceptional points, and complex-spectral topology
\cite{Bender1998_PRL,Bender2007_RPP,Rotter2009_JPA,Heiss2012_JPA,ElGanainy2018_NatPhys,MiriAlu2019_Science,Ozdemir2019_NatMat,
Ruter2010_NatPhys,Regensburger2012_Nature,Hodaei2017_Nature,Feng2017_NatPhoton,Bergholtz2021_RMP,Ashida2020_AdvPhys}.
In lattice systems, non-Hermiticity can strongly reshape localization and boundary responses, enabling boundary-state
accumulation, amplification, and wave manipulation beyond Hermitian paradigms
\cite{NHSE_review_Zhang2022,NonHermitianReview_Kawabata2020,1Manna20236}.

A paradigm of boundary sensitivity is the non-Hermitian skin effect (NHSE). Under open boundary conditions (OBC), 
a macroscopic number of eigenstates can become exponentially localized near a boundary, in stark contrast to the 
Bloch-like states of the same model under periodic boundary conditions (PBC). This breakdown of conventional Bloch 
band theory has motivated generalized bulk--boundary correspondences based on non-Bloch band theory and point-gap 
topology
\cite{YaoWang2018_PRL,Kunst2018_PRL,Gong2018_PRX,YokomizoMurakami2019_PRL,Kawabata2019_PRX,Okuma2020_PRL,Lee2016_PRL,Borgnia2020_PRL124056802},
and the resulting boundary spectra and signatures have been extensively investigated
\cite{Bergholtz2021_RMP,Ashida2020_AdvPhys,NHSE_review_Zhang2022,NonHermitianReview_Kawabata2020}.

Beyond uniform nonreciprocity, imaginary gauge fields offer a distinct route to non-Hermitian localization.
In the Hatano-Nelson model, a uniform imaginary vector potential modifies localization and induces strong boundary
sensitivity \cite{HatanoNelson1996_PRL,HatanoNelson1997_PRB,PRL119137402,PRL119137402,PRB103064201}.
More recently, Longhi showed that skin-like localization can arise purely from spatial fluctuations of the imaginary
gauge field even when its spatial average vanishes, so that the lattice remains globally reciprocal \cite{ENHSE_PRL2025}.
In this erratic non-Hermitian skin effect (ENHSE), eigenstates acquire an effective random envelope, and their
localization centers are controlled by extrema of the cumulative gauge field. This mechanism differs from the
conventional NHSE driven by net nonreciprocity and from Anderson localization.
A natural next question is how this envelope reshaping affects spectrally isolated in-gap boundary modes that are
already exponentially localized.

A natural arena for addressing this question is provided by quasiperiodic lattices, where localization and spectral gaps
emerge without uncorrelated disorder. The Aubry-Andr\'e-Harper (AAH) model is a canonical example, featuring a self-dual
localization transition \cite{AubryAndre1980} and close connections to the Harper equation and the Hofstadter problem
\cite{Harper1955,Hofstadter1976,Kohmoto1983_PRL}. Under OBC, bulk gaps can host spectrally isolated in-gap modes
localized near system boundaries, with their existence and profiles depending on the quasiperiodic phase offset
\cite{Thouless1983_Pump,Kraus2012_PRL,Kraus2013_PRL}. Quasiperiodic localization and gap physics have been realized in
photonic and cold-atom platforms \cite{Celi2014_PRL,Lahini2009_PRL,Verbin2013_PRL,Roati2008_Nature,Schreiber2015_Science}.
Non-Hermitian extensions of AAH-type models have also been studied, showing modified localization and anomalous boundary
or skin modes
\cite{Ganeshan2015_PRL,Longhi_NHAAH_PhaseTransitions,Zeng2020_NHAAH_Topo,Cai2021_NHAAH_pwave,Li2023_Entanglement_NHAAH,LocalizationTopo_NHQuasi}.
What remains unclear is how these spectrally isolated in-gap boundary modes respond when the hopping asymmetry is
generated by a zero-mean but strongly fluctuating imaginary gauge field (the erratic-gauge regime).

In this work, we address this question by employing an exact nonunitary gauge transformation under OBC.
This transformation maps the non-Hermitian chain onto its Hermitian AAH counterpart, so the open-boundary spectrum and
Lyapunov exponents are unchanged, while right/left eigenfunctions are reshaped by a spatial envelope set by the gauge field.
Within spectral gaps, we identify two isolated in-gap modes with sharply different responses to the gauge field: one remains pinned to the boundary, whereas the other stays exponentially localized but becomes gauge steerable, with its probability peak
repositioned by varying the gauge field without shifting its eigenvalue. We further propose a preparation protocol
based on weak local gain: the gain is applied at a single site selected from the Hermitian counterpart mode profile and then
kept fixed, which allows the steerable in-gap mode to be amplified from a bulk wave packet and, by changing the gauge field,
to be prepared at different locations.

The remainder of the paper is organized as follows. In Sec.~\ref{sec:model}, the continuum and
tight-binding models with Bernoulli gauge disorder is introduced. In Sec.~\ref{sec:Gauge mapping and Lyapunov exponents},
the nonunitary gauge transformation is derived under OBC and used to analyze the Lyapunov exponents
and the phase diagram is presented. In Sec.~\ref{sec:edge_ENHSE}, we show the coexistence of pinned
and steerable in-gap modes and verify this by finite-size adjudication. In Sec.~\ref{sec:dynamics},
we propose a weak local-gain protocol and demonstrate the dynamical preparation of the gauge-steerable
mode. Finally, we conclude in Sec.~\ref{sec:conclusions}.

\section{Model}
\label{sec:model}
A particle of mass $m$ moving in a one-dimensional bichromatic lattice subjected to an imaginary gauge field
can be described by the continuum Hamiltonian
\begin{equation}
H_{\rm cont}=\frac{\bigl(p-i A(x)\bigr)^2}{2m}
+V_1\sin^2(k_1 x)+V_2\cos\!\bigl(2k_2 x+\varphi\bigr),
\label{eq:Hcont}
\end{equation}
where $p=-i\partial_x$ is the momentum operator ($\hbar = 1$, hereafter), $A(x)\in\mathbb{R}$  is a random
non-Hermitian external field~\cite{HatanoNelson1996_PRL,HatanoNelson1997_PRB,PRL119137402,PRL119137402,PRB103064201},
$V_{1,2}$ are the lattice depths, and $k_{1,2}$ are the wave numbers of the primary and secondary lattices,
respectively. The phase $\varphi$ sets the spatial offset of the secondary lattice. We focus on a chain with
open boundaries, introducing $X(x)=\int_{x_0}^{x}A(x')\,dx',$ so that $X'(x)=A(x)$. Using $e^{-X}pe^{X}=p-i X'(x)$,
one obtains $e^{X}(p-i A)e^{-X}=p$ and hence the nonunitary gauge-transformed Hamiltonian reads
\begin{align}
H_{\rm cont}^{H}
&\equiv e^{X(x)} H_{\rm cont} e^{-X(x)} \notag\\
&= \frac{p^2}{2m}+V_1\sin^2(k_1 x)+V_2\cos\!\bigl(2k_2 x+\varphi\bigr).
\label{eq:HcontH}
\end{align}

In the deep-lattice regime $V_1\gg E_R$ (with recoil energy $E_R= k_1^2/2m$) and for a weaker secondary lattice
$V_1\gg V_2$, we project $H_{\rm cont}^{H}$ to the lowest Bloch band of the primary lattice and retain only onsite and
nearest-neighbor matrix elements. Let $a\equiv \pi/k_1$ be the primary-lattice spacing and $x_n\equiv na$ the lattice sites.
Expanding the field operator in lowest-band Wannier functions $w(x-x_n)$,
the resulting tight-binding Hamiltonian takes the Hermitian AAH form
\begin{equation}
\hat H^{H}=\sum_{n=1}^{N}V_n\,\hat c_n^\dagger \hat c_n
+\sum_{n=1}^{N-1}J\bigl(\hat c_{n+1}^\dagger \hat c_n+\hat c_n^\dagger \hat c_{n+1}\bigr),
\label{eq:H_H_AAH}
\end{equation}
where $\hat c_n$ annihilates a particle on site $n$ and $N$ is the number of lattice sites.
The nearest-neighbor hopping $J$ is the corresponding Wannier matrix element of $H_{\rm cont}^{H}$;
adopting the convention $J>0$, we define $J\equiv -\int \mathrm{d}x\, w(x-x_{n+1})\,H_{\rm cont}^{H}\,w(x-x_n),$
which is site-independent in the bulk for a uniform primary lattice.
\begin{figure}[t]
    \centering
    \includegraphics[width=0.8\linewidth]{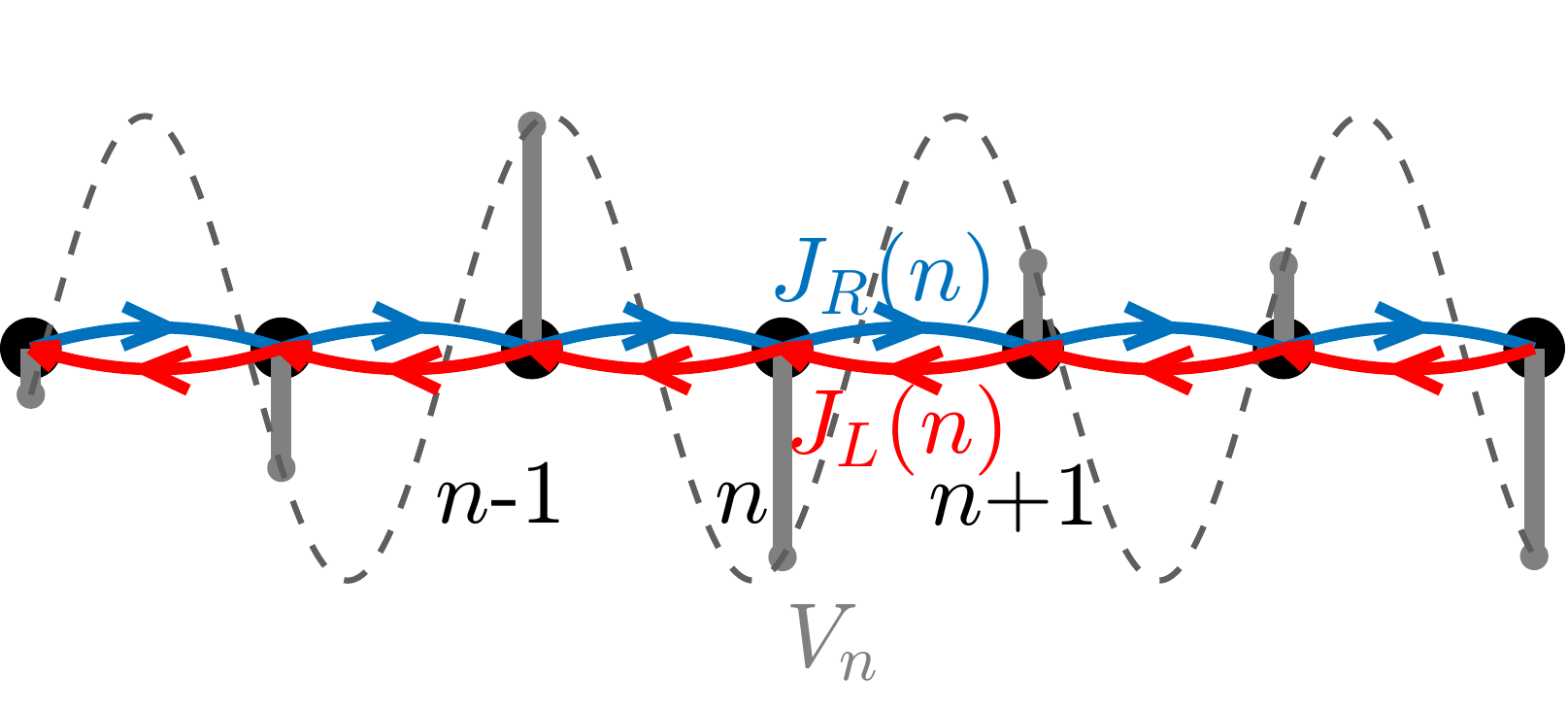}
    \caption{Schematic of the AAH model with disordered imaginary gauge field. Blue (red) arrows
    indicate $J_R(n)$ [$J_L(n)$]. The gray bars depict the onsite potential $V_n$ on each site,
    and the dashed curve is a guide for the quasiperiodic modulation profile.}
    \label{Model}
\end{figure}
The onsite potential $V_n$ is induced predominantly by the secondary lattice and is quasiperiodic,
\begin{equation}
V_n = \lambda \cos(2\pi \beta n + \varphi),
\end{equation}
where the incommensurability parameter is $\beta\equiv k_2/k_1$.
Within the same projection, using the localization of $w(x)$ (lowest band) and the even parity of $|w(x)|^2$,
the modulation amplitude can be written as $\lambda = V_2\int dx\,|w(x)|^2\cos(2k_2 x),$
with $w(x)$ the lowest-band Wannier function of the primary lattice centered at the origin.
In numerics we take $\beta$ as the golden ratio.

Finally, within the tight-binding description, we restore the imaginary gauge field of the original
continuum model in Eq.~\eqref{eq:Hcont} by applying the inverse nonunitary gauge
transformation to $\hat H^{H}$ with $X_n\equiv X(x_n)$. We define the transformation operator
$S=\exp\!\bigl(\sum_{n=1}^{N}X_n \hat c_n^\dagger \hat c_n\bigr)$ and obtain the non-Hermitian
tight-binding Hamiltonian by the nonunitary transformation,
\begin{align}
\hat H &= S \hat H^{H} S^{-1} \notag\\
&= \sum_{n=1}^{N} V_n\, \hat c_n^\dagger \hat c_n
+ \sum_{n=1}^{N-1}
\!\left[
J_R(n)\, \hat c_{n+1}^\dagger \hat c_n
+ J_L(n)\, \hat c_{n}^\dagger \hat c_{n+1}
\right].
\label{eq:H_model}
\end{align}
The right/left hopping amplitudes are
\begin{equation}
J_R(n) = J e^{h_n}, \qquad
J_L(n) = J e^{-h_n},
\label{eq:JRJL_def}
\end{equation}
where the discrete gauge increment is
\begin{equation}
h_n = X_{n+1}-X_n,\qquad X_1=0.
\label{eq:nsr}
\end{equation}
In what follows we focus on a Bernoulli gauge field,
\begin{equation}
h_n = s_n \Delta,
\label{eqbernouli}
\end{equation}
where $\{s_n\} = \pm1$ is a zero-mean Bernoulli sequence and $\Delta$ controls the disorder strength~\cite{ENHSE_PRL2025}.
This means the random non-Hermitian external field obeys Bernoulli distribution. The structure of the non-Hermitian
AAH chain considered here is schematically illustrated in Fig.~\ref{Model}.

\section{Nonunitary gauge transformation and phase structure}
\label{sec:Gauge mapping and Lyapunov exponents}
\subsection{Nonunitary gauge transformation}

To obtain an exact and analytically tractable description of the phase structure and in-gap modes in the presence of an erratic imaginary gauge field, we map the non-Hermitian Hamiltonian under OBC to its Hermitian AAH counterpart via a nonunitary gauge transformation, so that eigenvalues are preserved and non-Hermitian effects enter only through the spatial rescaling of eigenstates.
In the single-particle scenario, an eigenstate of
Hamiltonian~\eqref{eq:H_model} can be written as $|\psi\rangle=\sum_{n=1}^N \psi_n \hat c_n^\dagger|0\rangle$.
The coefficients $\psi_n$ and the corresponding eigenenergy $E$ satisfy the
stationary Schr\"odinger equation $\hat H|\psi\rangle=E|\psi\rangle$, which in
the site basis gives
\begin{align}
J_R(n-1)\psi_{n-1}+V_n\psi_n+J_L(n)\psi_{n+1}=E\psi_n .
\label{eq:eig_equation}
\end{align}
Using Eqs.~\eqref{eq:JRJL_def} and \eqref{eq:nsr}, the hopping asymmetry is gauged
 away by the rescaling
\begin{align}
\psi_n = e^{X_n}\,\phi_n .
\label{eq:wavefun_relation}
\end{align}
Substituting Eq.~\eqref{eq:wavefun_relation} into
Eq.~\eqref{eq:eig_equation} shows that the transformed amplitudes
$\phi_n$ obey a Hermitian AAH equation with symmetric
nearest-neighbor hopping $J$ and on-site potential $V_n$,
\begin{align}
J\,\phi_{n-1} + V_n\phi_n + J\,\phi_{n+1} = E\,\phi_n.
\label{eq:Herm_AAH_phi_short}
\end{align}
Thus the entire OBC spectrum of $\hat H$ coincides with the Hermitian counterpart AAH chain.
To characterize the asymptotic spatial growth/decay of a given eigenstate in the thermodynamic
limit, we introduce the Lyapunov exponent (spatial exponent)
\begin{align}
\Gamma(E)
= \left|\lim_{n\to\infty}\frac{1}{n}\ln\left|\frac{\psi_n}{\psi_1}\right|\right|.
\label{LyapunovI}
\end{align}
Using Eqs.~\eqref{eq:nsr} and \eqref{eq:wavefun_relation}, one obtains the exact decomposition~\cite{ENHSE_PRL2025}
\begin{align}
\Gamma(E)
= \tilde\Gamma(E) + \bar h ,
\label{eq:Lyap_relation_short}
\end{align}
where $\tilde\Gamma(E)$ is the corresponding exponent extracted from $\phi_n$, and
\begin{align}
\bar h = \lim_{n\to\infty}\frac{X_n}{n}
\end{align}
is the spatial average of the imaginary gauge field.
For the Bernoulli gauge disorder defined by Eq. \ref{eqbernouli}, $\bar h=0$ in the long-chain limit, so that
\begin{align}
\Gamma(E) = \tilde\Gamma(E).
\label{eq:Lyap_equal_short}
\end{align}
When an eigenenergy $E$ belonging to the continuous bulk spectrum of the Hermitian AAH chain,
$\tilde\Gamma(E)$ is known exactly and is independent of $E$~\cite{AubryAndre1980}:
it vanishes for $\lambda\le 2J$ and equals $\ln(\lambda/2J)$ for $\lambda>2J$.
Combining this result with Eq.~\eqref{eq:Lyap_equal_short}, we obtain the Lyapunov exponent
for bulk eigenstates of $\hat H$,
\begin{align}
\Gamma(E) =
\begin{cases}
0, & \lambda \le 2J,\\[4pt]
\ln\!\left(\dfrac{\lambda}{2J}\right), & \lambda>2J,
\end{cases}
\label{eq:NH_AAH_Lyap_short}
\end{align}
which yields the bulk localization length $\xi=1/\Gamma(E)$ for $\lambda>2J$.

For OBC eigenvalues $E$ that fall inside a bulk gap, the associated transfer matrix
has a nonzero Lyapunov exponent, implying an exponentially decaying tail whose decay rate is
energy dependent \cite{AubryAndre1980,Kohmoto1983_PRL,Thouless1983_Pump}. Nevertheless, for any fixed
eigenvalue $E$ the nonunitary gauge transformation implies that the spatial exponent extracted from $\psi_n$
differs from that of $\phi_n$ only through $\bar h$; hence for typical zero-mean gauge fields
with $\bar h=0$, the decay exponent of an in-gap state is inherited from the Hermitian counterpart.

\subsection{ENHSE-induced collective pinning and phase diagram}
\begin{figure}[t]
    \centering
    \includegraphics[width=\linewidth]{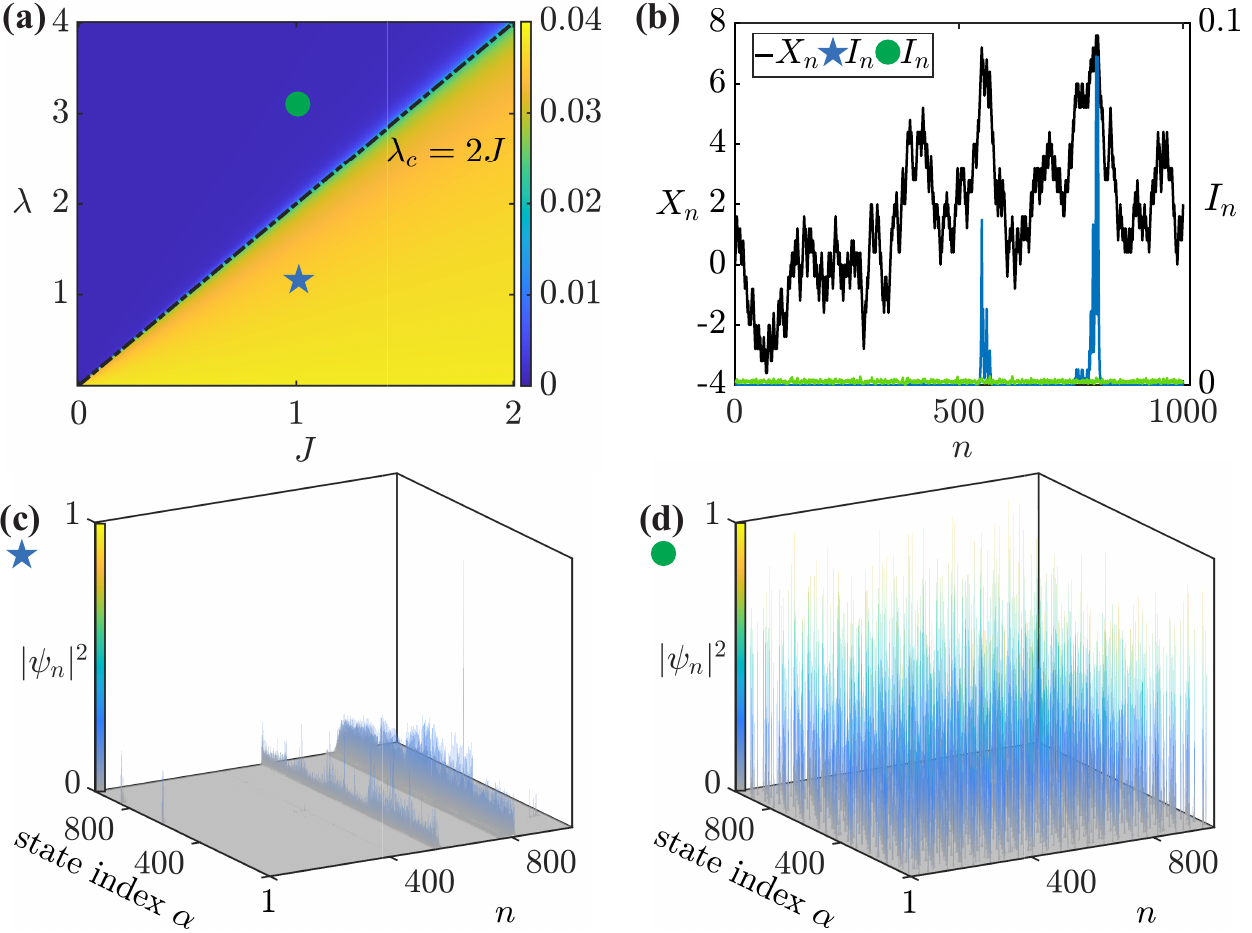}
    \caption{Phase structure and distribution of eigenstates when $\Delta=0.4$, $\beta=(\sqrt{5}-1)/2$
    and $\varphi=\pi/5$ hereafter for calculations. (a) Phase diagram indicated by $S_I$ in Eq.~\ref{eq:SI_def}
    versus $J$ and $\lambda$ when $N=1000$. $\lambda_c=2J$ is the demarcation line for different phases.
    The blue star and green circle indicate the parameter sets $(J,\lambda)=(1, 1)$ and $(1, 3)$, respectively.
    (b) $X_n$ defined in Eq.~\ref{eq:nsr} (black, left axis) and $I_n$ defined in Eq.~\ref{eq:In_def} (colored
    curves, right axis) for the two parameter sets marked in (a). (c), (d) Eigenstate density in the ENHSE
    regime $(J,\lambda)=(1, 1)$ and in the AAH-localized regime $(J,\lambda)=(1, 3)$, respectively.}
    \label{fig:fig1}
\end{figure}
To quantify the collective real-space structure of the spectrum of the non-Hermitian
Hamiltonian~\eqref{eq:H_model}, we introduce a statistical quantity for a site $n$
involving all eigenstates. Denoting the $N$ eigenstates by $\{\psi_{\alpha,n}\}$ with
eigenstate index $\alpha=1,\dots,N$, we define
\begin{align}
I_n=\frac{1}{N}\sum_{\alpha=1}^N |\psi_{\alpha,n}|^2 ,
\label{eq:In_def}
\end{align}
which measures the eigenstate-averaged weight at site $n$ (i.e., the average of
$|\psi_{\alpha,n}|^2$ over all right eigenstates). A quantity capturing the
global non-uniformity of this profile is
\begin{align}
S_I=\sum_{n=1}^N I_n^2 ,
\label{eq:SI_def}
\end{align}
which approaches its minimum value $S_I\simeq 1/N$ for a nearly uniform $I_n$, and increases
when major of the eigenstates with distribution concentrates near a few preferred sites.
Figure~\ref{fig:fig1}(a) shows the resulting phase diagram indicated by $S_I$ in the $(J,\lambda)$
plane (with $N=1000$). The dashed line $\lambda_c=2J$ marks the self-dual transition inherited from
the Hermitian AAH counterpart, across which the Lyapunov exponent in
Eq.~\eqref{eq:NH_AAH_Lyap_short} changes from $\Gamma=0$ to $\Gamma>0$.
In the regime $\lambda<2J$, the Hermitian counterpart supports extended bulk
eigenstates, yet the nonunitary envelope generated by the erratic imaginary
gauge field can collectively concentrate the physical right eigenstates
$\psi_{\alpha,n}$, yielding a strongly nonuniform $I_n$ and hence a large $S_I$.
In contrast, for $\lambda>2J$ the quasiperiodic potential already localizes the
eigenstates of the Hermitian counterpart, and the spectrum-averaged intensity becomes
much less sensitive to the envelope of $X_n$, resulting in a significantly
smaller $S_I$.

The statistical details of the distinct behavior are illustrated in Fig.~\ref{fig:fig1}(b)-(d)
for the two representative parameter sets marked in Fig.~\ref{fig:fig1}(a): $(J,\lambda)=(1,1)$ (blue star, $\lambda<2J$) and
$(J,\lambda)=(1,3)$ (green circle, $\lambda>2J$).
Figure~\ref{fig:fig1}(b) directly compares, for each parameter set, the cumulative
profile $X_n$ (black curve, left axis) and $I_n$ defined in Eq.~\eqref{eq:In_def} (colored curve, right axis).
For $(J,\lambda)=(1,1)$, where the eigenstates of the Hermitian counterpart are
extended, the nonunitary envelope $e^{X_n}$ selects only a few preferred sites;
accordingly, $I_n$ in Fig.~\ref{fig:fig1}(b) develops sharp peaks located near the
dominant maxima of $X_n$. This collective pinning is further corroborated by the
eigenstate-density shown in Fig.~\ref{fig:fig1}(c), where a large fraction of the
eigenstates accumulates its weight around the same spatial locations.
For $(J,\lambda)=(1,3)$, by contrast, the Hermitian counterpart is in the AAH-localized
regime, so individual eigenstates are exponentially localized with centers distributed
across the chain. As a result, $I_n$ in Fig.~\ref{fig:fig1}(b) does not display pronounced
peaks but stays close to a weakly modulated background, showing no systematic correlation
with the extreme of $X_n$. The eigenstate-density in Fig.~\ref{fig:fig1}(d) confirms that
the states remain exponentially localized with distinct localization centers, consistent
with the much smaller value of $S_I$ in this region.

\section{Coexistence of boundary-pinned and gauge-steerable edge states}
\label{sec:edge_ENHSE}
\begin{figure}[t]
    \centering
    \includegraphics[width=\linewidth]{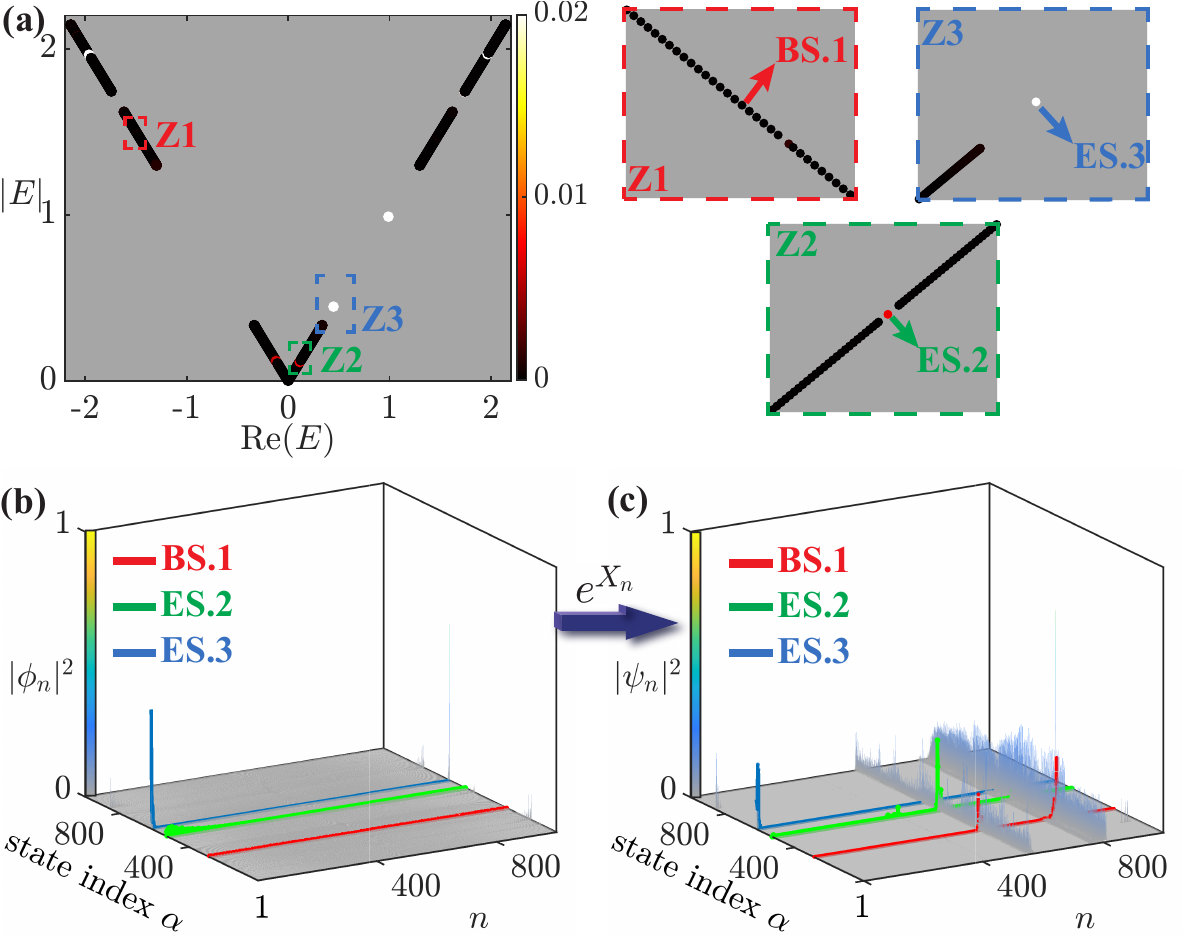}
    \caption{Spectrum and structures for the eigenstates at the parameter set indicated by the
    blue star marker, i.e., $(J, \lambda)=(1, 1)$, in Fig.~\ref{fig:fig1}. (a) Spectrum of
    $(\mathrm{Re}\, E, |E|)$, with color representing the Lyapunov exponent (\ref{LyapunovI})
    in open boundary condition. Zoomed windows Z1, Z2, and Z3 mark the bulk state BS.1 and the
    in-gap states ES.2 and ES.3. In (b) and (c), eigenstates are ordered by increasing $\mathrm{Re}\,E$.
    (b) Distribution of the eigenstates for the Hermitian AAH Hamiltonian (\ref{eq:H_H_AAH})
    versus the eigenstate index $\alpha$ and the site index $n$. (c) Corresponding distribution
    of the eigenstates for the ENHSE-AAH Hamiltonian (\ref{eq:H_model}) with the same eigenvalues.
    In (b) and (c) the states BS.1, ES.2, and ES.3 are highlighted. Each eigenstate is normalized.}
    \label{fig:edge_profiles}
\end{figure}
With the nonunitary gauge transformation shown above, we can analyze boundary physics using the Hermitian
AAH spectrum as a reference: the eigen-spectrum is maintained in OBC, and for $\bar{h}_n = 0$ the spatial
decay/growth exponents are inherited from the Hermitian counterpart. In the ENHSE regime $\lambda<2J$,
bulk eigenstates are extended in the Hermitian counterpart, yet there still exist exponentially decaying
eigenstates with eigenvalues lying inside the bulk gaps~\cite{AubryAndre1980}, which can be seen in
Fig.~\ref{fig:edge_profiles}. The gauge transformation retains the corresponding in-gap eigenvalues
while reshaping the envelopes of the corresponding eigenstates through $e^{X_n}$. Figure~\ref{fig:edge_profiles}(a) shows the OBC spectrum of $\hat H$, from which we select one representative bulk state (BS.1) and two representative in-gap modes (ES.2 and ES.3), whose Hermitian counterparts are edge states for the comparison below.

Figs.~\ref{fig:edge_profiles}(b) and \ref{fig:edge_profiles}(c) compare the
real-space distribution of all eigenstates before and after the nonunitary transformation in
Eq.~(\ref{eq:H_model}). As shown in Fig.~\ref{fig:edge_profiles}(b), the amplitudes $|\phi_n|^2$
of BS.1 is extended along the chain, while those of ES.2 and ES.3 are peaked near the left
boundary. Fig.~\ref{fig:edge_profiles}(c) shows the corresponding ENHSE-AAH amplitudes $|\psi_n|^2$,
obtained corresponding to the same eigenvalues via $\psi_n = e^{X_n}\phi_n$ for one Bernoulli
sequence. The envelope $e^{X_n}$ leaves all eigenvalues and Lyapunov exponents unchanged but
reshapes the spatial profiles. For the bulk state BS.1, the transformation produces the familiar
ENHSE collective pinning: the state of the Hermitian counterpart behaves extendedly in
Fig.~\ref{fig:edge_profiles}(b), while its distribution in Fig.~\ref{fig:edge_profiles}(c) is
concentrated near the global maximum of $X_n$. The two in-gap modes ES.2 and ES.3 respond in
a qualitatively different manner. ES.3 remains localized at the edge before and after the
transformation, i.e., its main peak is always pinned to the left boundary and only its tail is
slightly distorted by the envelope. In contrast, ES.2 in Fig.~\ref{fig:edge_profiles}(c) is
pulled away from the left edge and re-centered at an interior position within the bulk, where
$X_n$ develops a local peak.

\begin{figure}[t]
    \centering
    \includegraphics[width=\linewidth]{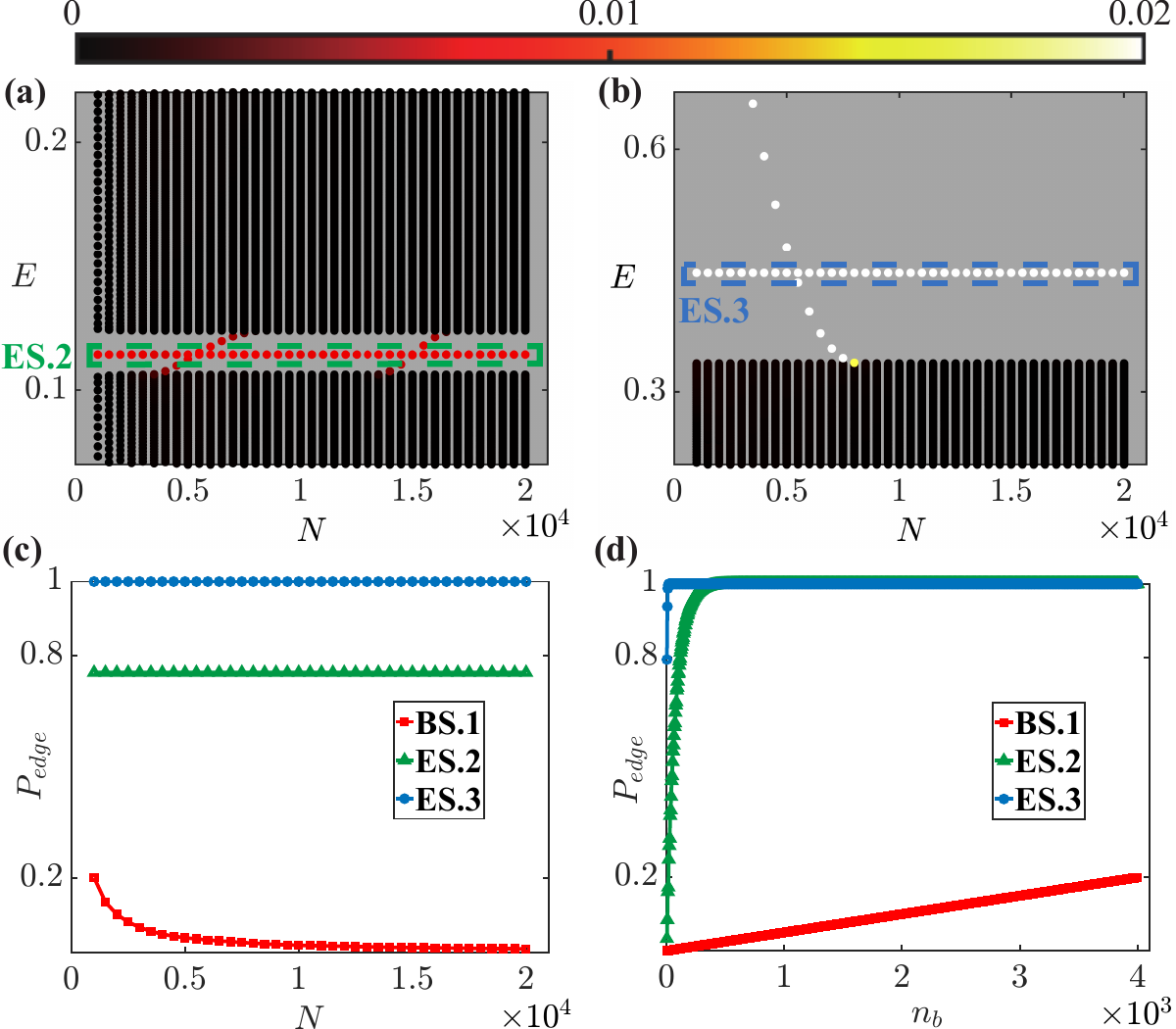}
    \caption{Finite-size adjudication of the three eigenstates BS.1, ES.2 and ES.3 for the Hermitian
    counterpart when $J=\lambda=1$, $\beta=(\sqrt{5}-1)/2$, and $\varphi=\pi/5$.
(a) Eigenenergies in the gap hosting ES.2 versus the system size $N$. (b) Same as (a) but for the
gap hosting ES.3. (c) Boundary weight $P_{\mathrm{edge}}$ defined in Eq.~(\ref{eq:Pedge_def}) of BS.1,
ES.2, and ES.3 versus $N$ evaluated with a fixed boundary window length $n_b=100$. (d) Boundary weight
$P_{\mathrm{edge}}$ of BS.1, ES.2, and ES.3 versus the boundary window size $n_b$ at $N=40000$.}
\label{fig:size_scaling}
\end{figure}

To establish that the Hermitian modes corresponding to ES.2 and ES.3 are bona fide edge states, i.e., spectrally
isolated and localized near the boundaries rather than merging into the bulk continua with the size increasing,
we perform finite-size adjudication for the Hermitian AAH chain. We start from a chain of size $N=1000$ and increase
$N$ up to $20000$ in steps of $\Delta N=500$, keeping $(J,\lambda,\beta,\varphi)$ fixed. We monitor how the in-gap
eigenvalues within the regions Z2 (ES.2) and Z3 (ES.3) in Fig.~\ref{fig:edge_profiles}(a) evolve versus the system
size $N$ in Figs.~\ref{fig:size_scaling}(a) and~\ref{fig:size_scaling}(b). It can be seen that the eigenvalues of ES.2
and ES.3 remain inside bulk gaps and do not merge into the neighboring continua as $N$ increases, indicating spectral
isolation in the thermodynamic limit.

To quantify the boundary localization of an eigenstate with amplitude $\phi_n$ at site $n$, we compute the boundary
weight
\begin{align}
P_{\mathrm{edge}}=\sum_{n=1}^{n_b}|\phi_n|^2+\sum_{n=N-n_b+1}^{N}|\phi_n|^2,
\label{eq:Pedge_def}
\end{align}
where $n_b$ is the window length measured from each boundary. The larger $P_{\mathrm{edge}}$ means that the eigenstate
is more concentrated at the edges of the chain at fixed $n_b$. As shown in Fig.~\ref{fig:size_scaling}(c) when $n_b = 100$,
the boundary weight $P_{\mathrm{edge}}$ of the bulk state BS.1 decreases as the system size $N$ increases. By contrast,
the $P_{\mathrm{edge}}$ values of ES.2 and ES.3 are independent of $N$. This means ES.2 and ES.3 are
boundary-localized, but BS.1 is not. This is further supported by Fig.~\ref{fig:size_scaling}(d): at fixed $N=40000$,
the boundary weights of ES.2 and ES.3 rapidly approach unity as $n_b$ increases, whereas BS.1 remains far
from saturation. Together, these finite-size adjudications establish that the Hermitian counterparts of ES.2 and ES.3
are bona fide edge states. The nonunitary gauge transformation in~Eq.(\ref{eq:H_model}) then guarantees that the
corresponding OBC eigenvalues are unchanged, while the associated right/left eigenfunctions are reshaped by $X_n$,
which can shift the probability peak of ES.2 without altering its in-gap eigenvalue.

\begin{figure}[t]
    \centering
    \includegraphics[width=\linewidth]{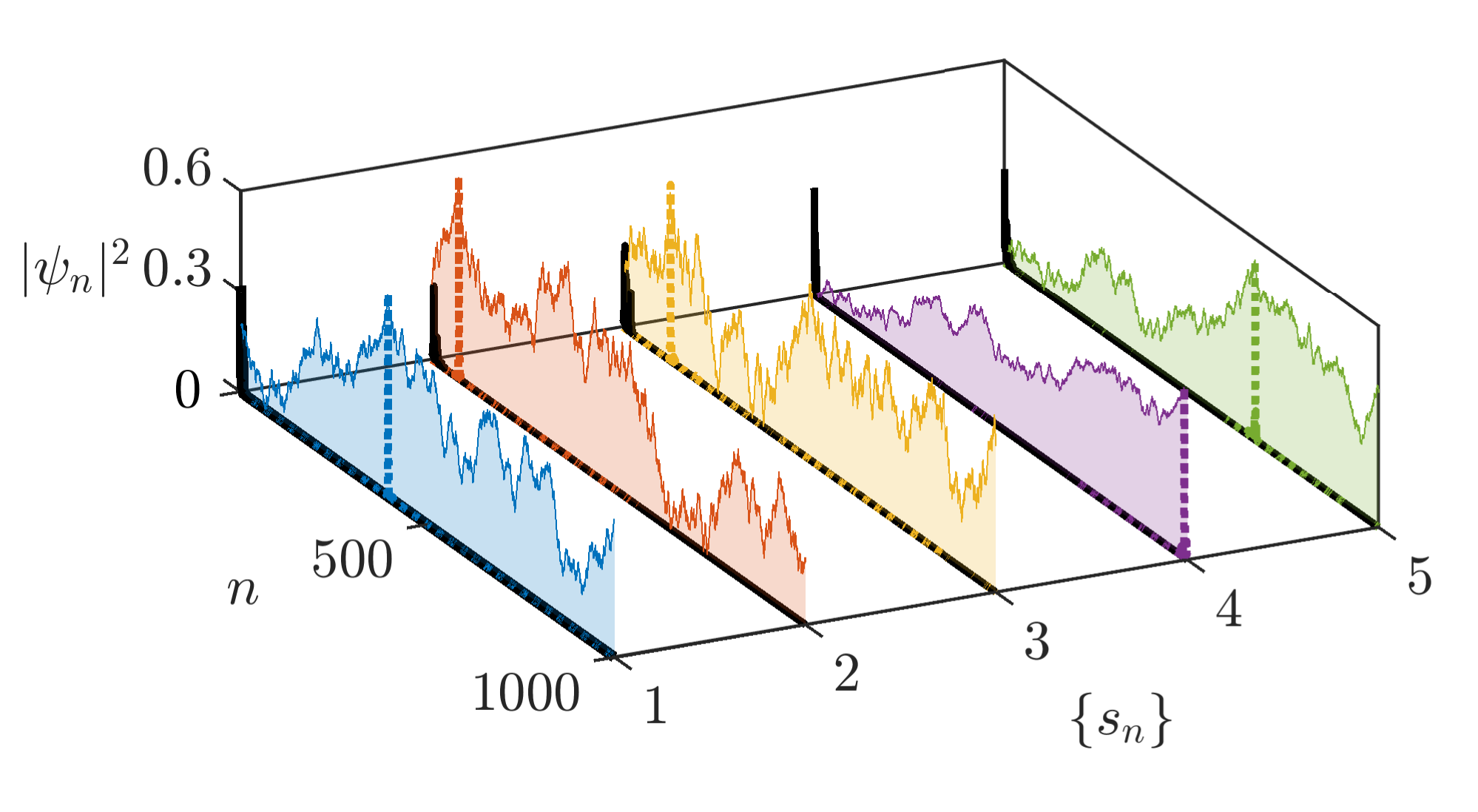}
    \caption{Distributions of the eigenstates ES.2 and ES.3 for five different Bernoulli sequences
    $\{s_n\}$. All parameters are identical to those used for ES.2 and ES.3 in
    Fig.~\ref{fig:edge_profiles}, only the Bernoulli sequence $\{s_n\}$ varies.
    Colored dashed curves represent $|\psi_n|^2$ of ES.2, black solid curves represent that of ES.3;
    the corresponding $X_n$ are plotted as colored solid curves with semi-transparent filled areas.}
    \label{fig:5sn}
\end{figure}

To check the distinct responses of ES.2 and ES.3 to different $X_n$, we keep $(J,\lambda,\beta,\varphi)$ fixed and
vary only the Bernoulli sequence $\{s_n\}$. Fig.~\ref{fig:5sn} shows representative densities
$|\psi_n|^2$ of ES.2 and ES.3 for five independent $\{s_n\}$. For ES.2, all curves exhibit
comparable exponential tails, yet their peaks shift over the chain, tracking prominent peaks of the cumulative gauge
profile $X_n$. In contrast, the edge mode ES.3 remains essentially pinned to the boundary. Accordingly, we refer to
ES.2 as a gauge-steerable in-gap mode, i.e., its localization center can be repositioned by changing the imaginary-gauge
configuration $A(x)$, whereas ES.3 pinned tenaciously at the edge, insensitive to the gauge randomness.

\section{Local-gain protocol for preparing the gauge-steerable in-gap state}
\label{sec:dynamics}

To go beyond a static eigenstate analysis, we aim to prepare the gauge-steerable in-gap mode ES.2 from a bulk wave packet using a weak local-gain protocol. Without this gain, the state retains substantial overlap with many bulk modes, so ES.2 is not preferentially selected. We show that weak on-site gain at a single site provides a mode selector: for an isolated eigenmode it induces a small imaginary-energy shift, and the mode with the largest positive shift dominates the long-time normalized dynamics.

Let $\{|\psi^{\mathrm{R}}_\alpha\rangle\}$ and $\{|\psi^{\mathrm{L}}_\alpha\rangle\}$
be the right and left eigenstates of $\hat H$,
\begin{align}
\hat H|\psi^{\mathrm{R}}_\alpha\rangle
&= E_\alpha|\psi^{\mathrm{R}}_\alpha\rangle,\\
\hat H^\dagger|\psi^{\mathrm{L}}_\alpha\rangle
&= E_\alpha^*|\psi^{\mathrm{L}}_\alpha\rangle,
\end{align}
with biorthogonal normalization
$\langle\psi^{\mathrm{L}}_\alpha|\psi^{\mathrm{R}}_\beta\rangle=\delta_{\alpha\beta}$.
Consider a local on-site perturbation applied at lattice site $n$,
\begin{align}
\delta\hat V = \delta v\,|n\rangle\langle n|,
\end{align}
where $\delta v$ is a complex perturbation.
For an isolated, nondegenerate eigenvalue $E_\alpha$, first-order perturbation theory gives~\cite{Bai2025_SMT_SHE}
\begin{align}
\delta E_\alpha
= \langle\psi^{\mathrm{L}}_\alpha|\delta\hat V|\psi^{\mathrm{R}}_\alpha\rangle
= \delta v\,\chi_\alpha(n),
\label{eq:deltaE}
\end{align}
where the biorthogonal weight is
\begin{align}
\chi_\alpha(n)
= \langle\psi^{\mathrm{L}}_\alpha|n\rangle\langle n|\psi^{\mathrm{R}}_\alpha\rangle
= \psi^{\mathrm{L}*}_\alpha(n)\,\psi^{\mathrm{R}}_\alpha(n).
\label{eq:chi_general_def}
\end{align}
For such a perturbation, $\chi_\alpha(n)$ controls the first-order eigenvalue shift via Eq.~\eqref{eq:deltaE}.
We take $\delta v=i\gamma$, i.e., on-site gain, with $J\gg\gamma>0$, so that
$\mathrm{Im}\,\delta E_\alpha = \gamma\,\chi_\alpha(n)$ for an isolated mode. Thus, for a given target mode $\alpha$,
we choose
\begin{align}
n_0=\arg\max_n \chi_\alpha(n),
\label{eq:22}
\end{align}
which yields the maximal imaginary shift and this makes the target mode dominant in the long-time normalized dynamics.

Equation~\eqref{eq:wavefun_relation} allows one to evaluate $\chi_\alpha(n)$ from the Hermitian counterpart.
Let $\hat H^{H}$ be the Hermitian AAH Hamiltonian corresponding to $\hat H$ under OBC by the nonunitary gauge
transformation, and let $|\phi_\alpha\rangle$ be its normalized eigenstates,
\begin{align}
\hat H^{H}|\phi_\alpha\rangle = E_\alpha|\phi_\alpha\rangle,\qquad
\langle\phi_\alpha|\phi_\beta\rangle=\delta_{\alpha\beta}.
\end{align}
The right/left eigenstates of $\hat H$ are then
\begin{align}
|\psi^{\mathrm{R}}_\alpha\rangle = S|\phi_\alpha\rangle,\qquad
|\psi^{\mathrm{L}}_\alpha\rangle = (S^{-1})^\dagger|\phi_\alpha\rangle,
\label{eq:S_transform_RL_general}
\end{align}
with $S$ being diagonal in the site basis. It follows immediately that
\begin{align}
\chi_\alpha(n)=\psi^{\mathrm{L}*}_\alpha(n)\psi^{\mathrm{R}}_\alpha(n)=|\phi_\alpha(n)|^2.
\label{eq:chi_from_hermitian}
\end{align}
Therefore the optimal gain site is fixed by the profile of the Hermitian mode and is independent of the
erratic-gauge field.

\begin{figure}[t]
    \centering
    \includegraphics[width=\linewidth]{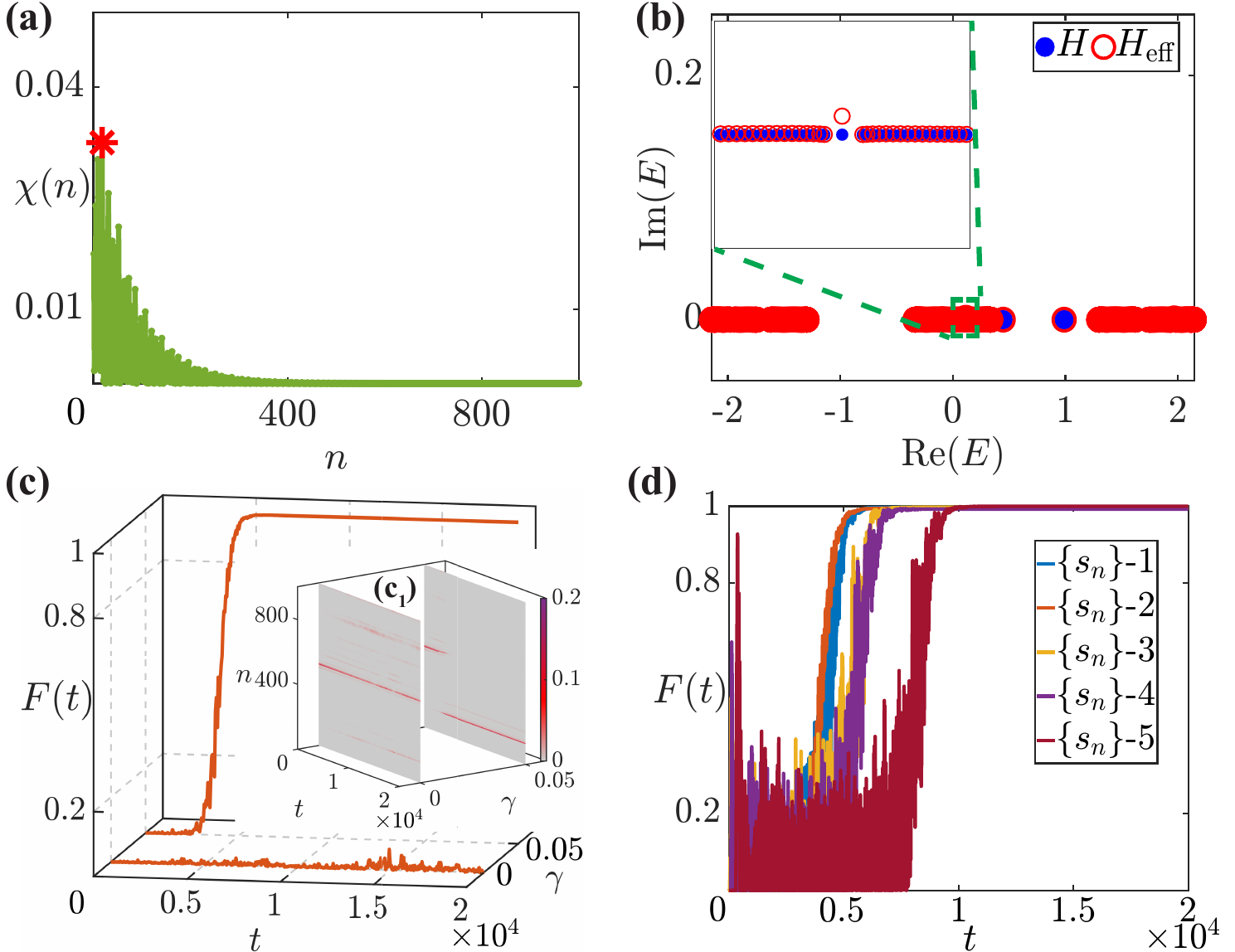}
    \caption{Local-gain protocol for dynamically preparing the gauge-steerable in-gap mode ES.2.
    (a) Biorthogonal weight $\chi(n)$ of ES.2 in Eq.~(\ref{chi29}); the star marks the selected gain
    site $n_0=18$. (b) Complex spectra for the unperturbed Hamiltonian $\hat H$ defined in
    Eq.~(\ref{eq:H_model})(blue dots) and the gain-perturbed Hamiltonian $\hat H_{\mathrm{eff}}$
    defined in Eq.~(\ref{eq:Heff_def}) (red open circles) for $\gamma=0.05$; the inset zooms into
    the gap hosting ES.2. (c) Fidelity $F(t)$ to ES.2 for the Bernoulli sequence $\{s_n\}$-2
    (same labeling as in Fig.~\ref{fig:5sn}), for $\gamma=0$ and $\gamma=0.05$. Inset (c$_1$):
    the evolution of the densities $|\Psi_n(t)|^2$ for $\gamma=0$ and $\gamma=0.05$. (d) $F(t)$
    for five Bernoulli sequences $\{s_n\}$-$j$ ($j=1,\dots,5$) used in Fig.~\ref{fig:5sn},
    evaluated when $\gamma=0.05$ with the same fixed gain site $n_0$.}
    \label{fig:dynamics}
\end{figure}

Based on the analysis above, we now implement the local gain to prepare the gauge-steerable in-gap mode
ES.2. Let $|\phi_{\mathrm{ES.2}}\rangle$ denote the corresponding Hermitian eigenstate (with eigenvalue
$E_{\mathrm{ES.2}}$) identified in Sec.~\ref{sec:edge_ENHSE}. From Eq.~\eqref{eq:chi_from_hermitian} we
obtain
\begin{align}
\chi_{\mathrm{ES.2}}(n)=|\phi_{\mathrm{ES.2}}(n)|^2.
\label{chi29}
\end{align}
The resulting $\chi_{\mathrm{ES.2}}(n)$ is shown in Fig.~\ref{fig:dynamics}(a), where it attains its
maximum at $n=18$; according to Eq.~\eqref{eq:22}, $n_0 = n$. With $n_0$ fixed, we introduce weak local
gain and define
\begin{align}
\hat H_{\mathrm{eff}}
= \hat H + i\gamma\,|n_0\rangle\langle n_0|.
\label{eq:Heff_def}
\end{align}
The gain term breaks the exact nonunitary gauge transformation in Eq.~\ref{eq:H_model}. As $\gamma \ll J$,
it produces only a small perturbation to the eigenstates, while shifting the eigenvalues according to the
biorthogonal weight. In particular, the targeted mode acquires the largest positive imaginary shift,
\begin{align}
\mathrm{Im}\,\delta E_{\mathrm{ES.2}} = \gamma\,\chi_{\mathrm{ES.2}}(n_0).
\end{align}
With this choice of $n_0$, ES.2 receives a parametrically larger positive imaginary shift than the
remaining modes, as shown in Fig.~\ref{fig:dynamics}(b), and therefore dominates the long-time
normalized dynamics.

The dynamics is governed by
\begin{align}
i\frac{d}{dt}|\Psi(t)\rangle
= \hat H_{\mathrm{eff}}|\Psi(t)\rangle .
\label{eq:Schrod_Heff}
\end{align}
We numerically solve Eq.~\eqref{eq:Schrod_Heff} with normalization of the state during the evolution.
A normalized Gaussian wave packet centered in the bulk is chosen as the initial state,
\begin{align}
\Psi_n(0)\propto
\exp\!\left[-\frac{(n-n_c)^2}{2\sigma^2}\right],\qquad
n_c=\frac{N+1}{2},
\end{align}
with $\sigma=20$. To confirm the preparation of ES.2, we monitor the fidelity with the right mode of
ES.2, $|\psi^{\mathrm{R}}_{\mathrm{ES.2}}\rangle$, versus time,
\begin{align}
F(t)
= \big|\langle\psi^{\mathrm{R}}_{\mathrm{ES.2}}|\Psi(t)\rangle\big|^2.
\label{eq:Fidelity_def}
\end{align}

Figure~\ref{fig:dynamics}(c) summarizes the dynamics for the Bernoulli sequence $\{s_n\}$-2
(the same labeling as in Fig.~\ref{fig:5sn}). In Fig.~\ref{fig:dynamics}(c$_1$), the two space-time density maps
compare $\gamma=0$ and $\gamma=0.05$. For $\gamma=0$ (no local gain), the wave packet spreads and retains
substantial bulk components rather than converging to ES.2. This is quantified in Fig.~\ref{fig:dynamics}(c):
the fidelity with ES.2 stays significantly below unity throughout the evolution. By contrast, when a weak local
gain is applied ($\gamma=0.05$), Fig.~\ref{fig:dynamics}(c$_1$) shows that the normalized profile is gradually
drawn from the bulk toward the left region and eventually converges to a sharply localized pattern whose maximum
coincides with the ENHSE-steered peak position of ES.2. Consistently, the fidelity in Fig.~\ref{fig:dynamics}(c)
increases from $F(0)\ll 1$ and saturates to unity, indicating that the local gain selectively enhances the ES.2
component.

To test the generality, we repeat the same protocol for the five independent Bernoulli sequences used in
Fig.~\ref{fig:5sn}. The gain site is kept fixed at the same $n_0$ determined once from the Hermitian counterpart,
while the long-time peak position follows the envelope $e^{X_n}$. As shown in Fig.~\ref{fig:dynamics}(d), all
fidelities saturate near unity, demonstrating that the preparation works consistently across different erratic-gauge
fields and yields a dynamically selected, position-tunable in-gap mode.

\section{Conclusion}
\label{sec:conclusions}
We have investigated a quasiperiodic non-Hermitian AAH chain with a zero-mean, spatially fluctuating imaginary
gauge field under OBC. An exact nonunitary gauge transformation mapped the system to its Hermitian AAH counterpart:
the spectrum and Lyapunov exponents were unchanged, while the right/left eigenstates were reshaped by the gauge-induced
envelope. In a parameter region with spectrally isolated in-gap edge modes, we identified an counterintuitive
in-gap state that remains exponentially localized but whose probability peak can be shifted by varying only the
imaginary gauge field, in contrast to a second mode that stays pinned to the boundary. We further showed that this steerable in-gap mode can be dynamically prepared from a bulk wave packet by applying weak on-site gain at a single fixed site, chosen where the corresponding Hermitian eigenstate attains its maximum probability density. With the gain site kept unchanged, varying the gauge field produces long-time states that remain exponentially localized but peak at different positions. These results provide a simple route to generating position-tunable, exponentially localized in-gap states in non-Hermitian quasiperiodic lattices.

\begin{acknowledgments}
X.L.Z. thanks Zhi Li and Peng Zou for helpful discussions. This work was supported by the Joint
Fund of Natural Science Foundation of Shandong Province (Grant No. ZR2024LLZ004), the National
Natural Science Foundation of China(Grant No. 12575010, No. 12005110), and the Natural Science
Foundation of Shandong Province (Grant No. ZR2020QA078, No. ZR2023QB065, No. ZR2023MD064).
\end{acknowledgments}


\end{document}